\documentclass[12pt]{article}
\usepackage{amsmath}
\usepackage{epsfig}
\newcommand{\ep}{\varepsilon}

\catcode`@=11
\newcount\@tempcntc
\def\@citex[#1]#2{\if@filesw\immediate\write\@auxout{\string\citation{#2}}\fi
  \@tempcnta\z@\@tempcntb\m@ne\def\@citea{}\@cite{\@for\@citeb:=#2\do
    {\@ifundefined
       {b@\@citeb}{\@citeo\@tempcntb\m@ne\@citea\def\@citea{,}{\bf ?}\@warning
       {Citation `\@citeb' on page \thepage \space undefined}}%
    {\setbox\z@\hbox{\global\@tempcntc0\csname b@\@citeb\endcsname\relax}%
     \ifnum\@tempcntc=\z@ \@citeo\@tempcntb\m@ne
       \@citea\def\@citea{,}\hbox{\csname b@\@citeb\endcsname}%
     \else
      \advance\@tempcntb\@ne
      \ifnum\@tempcntb=\@tempcntc
      \else\advance\@tempcntb\m@ne\@citeo
      \@tempcnta\@tempcntc\@tempcntb\@tempcntc\fi\fi}}\@citeo}{#1}}
\def\@citeo{\ifnum\@tempcnta>\@tempcntb\else\@citea\def\@citea{,}%
  \ifnum\@tempcnta=\@tempcntb\the\@tempcnta\else
   {\advance\@tempcnta\@ne\ifnum\@tempcnta=\@tempcntb \else \def\@citea{--}\fi
    \advance\@tempcnta\m@ne\the\@tempcnta\@citea\the\@tempcntb}\fi\fi}
\catcode`@=12
\begin{document}

%
%
\title{
\vskip-3cm{\baselineskip14pt
\centerline{\normalsize DESY~11--087\hfill ISSN~0418--9833}
\centerline{\normalsize May 2011\hfill}}
\vskip1.5cm
Counting master integrals: integration by parts vs.\ differential reduction
}

\author{
{\sc Mikhail~Yu.~Kalmykov}\thanks{On leave of absence from
Joint Institute for Nuclear Research, 141980 Dubna (Moscow Region), Russia.},
{\sc Bernd~A.~Kniehl}
\\
\\
{\normalsize II. Institut f\"ur Theoretische Physik, Universit\"at Hamburg,}\\
{\normalsize Luruper Chaussee 149, 22761 Hamburg, Germany}
}

\date{}

\maketitle
\abstract{
The techniques of integration by parts and differential reduction differ in the
counting of master integrals.
This is illustrated using as an example the two-loop sunset diagram with
on-shell kinematics.
A new algebraic relation between the master integrals of the two-loop sunset
diagram that does not follow from the standard integration-by-parts technique
is found. 
\medskip

\medskip

\noindent
Dedicated to K.G. Chetyrkin on the occasion of his sixtieth birthday.
}

\newpage


\renewcommand{\thefootnote}{\arabic{footnote}}
\setcounter{footnote}{0}

Within dimensional regularization \cite{dimreg}, 
the integration-by-parts (IBP) technique \cite{ibp} (for a recent review, see
Ref.~\cite{Grozin}) is one of the most powerful tools for evaluating multiloop
Feynman diagrams. 
However, the question how to construct algebraical, geometrical, or topological 
criteria for the (ir)reducibility of Feynman diagrams in general is still an
open one \cite{reducibility}. 
At present, only the explicit solution of recurrence relations (see, for
example, Ref.~\cite{Tarasov}), or the application of the Laporta algorithm
\cite{Laporta}, as implemented in computer programs \cite{Laporta:computer},
provide an answer on this question.

In a series of recent publications \cite{Kalmykov,BKK,review,Bytev:2011ks}, we
addressed the problem of counting nontrivial master integrals with the help of
hypergeometric representations for Feynman diagrams and the
differential-reduction algorithm \cite{Takayama}.
By detailed analysis, we thus derived the following empirical criteria for the
hypergeometric representation of Feynman diagrams:
If a Feynman diagram is expressible as a linear combination of Horn-type
hypergeometric functions with rational coefficients, then 
(i) each hypergeometric function has the same number of basis elements in the
framework of differential reduction \cite{Takayama}; 
and (ii) the number of nontrivial master integrals is equal to the number of
basis elements of the hypergeometric functions (up to the modules of Feynman
integrals expressible in terms of products of algebraic functions and
$\Gamma$ functions).

These criteria were conjectured on the basis of the analysis of a variety of
particular examples \cite{Kalmykov,BKK,review}.
A rigorous mathematical proof is still lacking.
The aim of this Letter is to present a first example where the application of
our criteria allow us to find an additional algebraic relation between master
integrals which does not follow from the standard IBP technique.
In fact, Tarasov's algorithm \cite{Tarasov} for the reduction of two-loop
propagator diagrams, as implemented in the computer packages of
Ref.~\cite{Mertig:1998vk}, and Laporta's algorithm \cite{Laporta}, as
implemented in the computer package of Ref.~\cite{Laporta:computer}, fail to
produce this relation.

\begin{figure}[th]
\centering
{\vbox{\epsfysize45mm \epsfbox{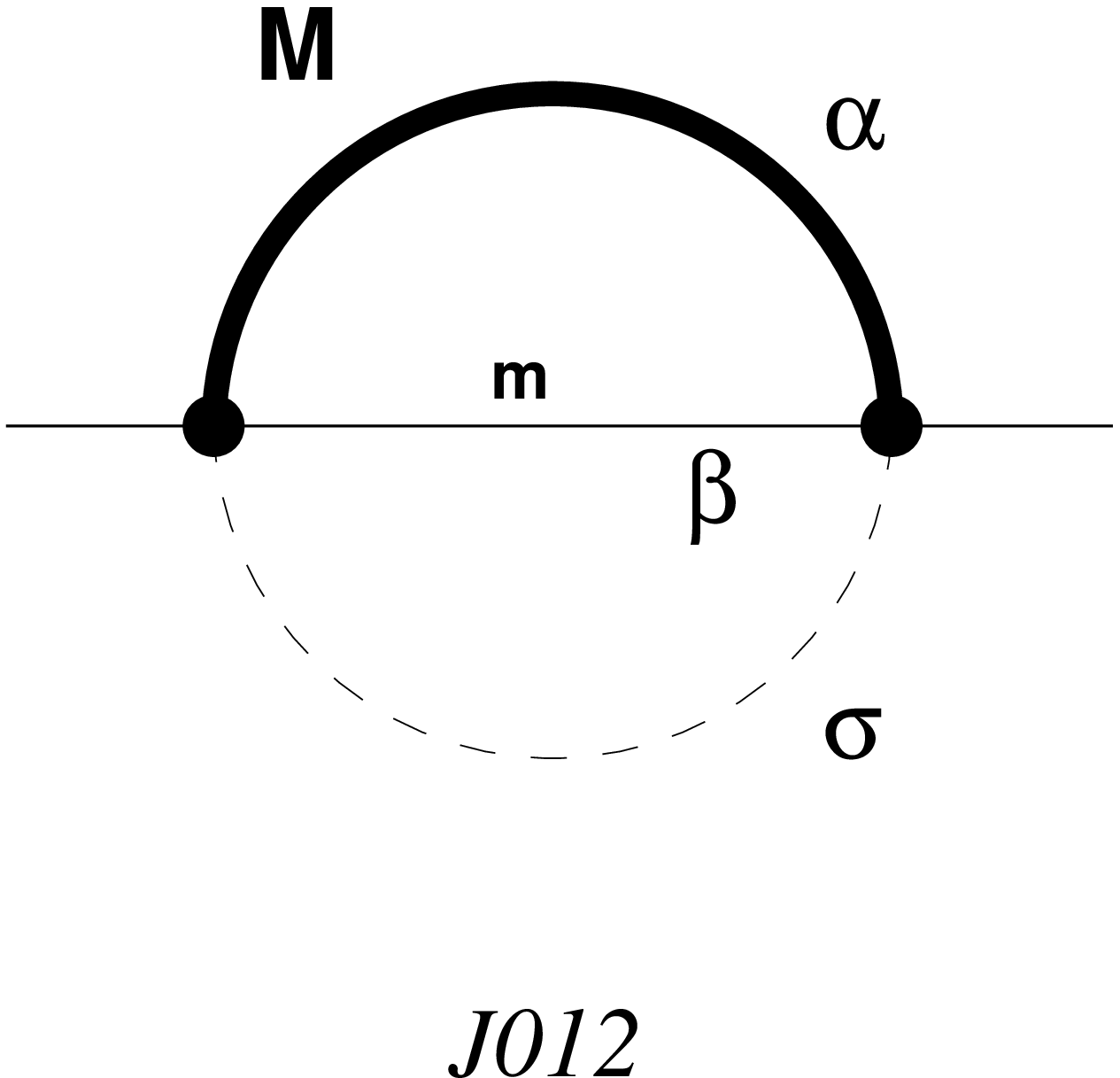}}}
\caption{
Two-loop self-energy sunset-type diagram $J_{012}$.}
\label{j012}
\end{figure}

Let us consider the two-loop self-energy sunset-type diagram $J_{012}$ with
on-shell kinematics, defined as
\begin{equation}
J_{012}(\sigma,\beta,\alpha) =\pi^{-n}
\int 
\left.
\frac{d^nk_1d^nk_2}{[(p-k_1)^2]^\sigma[(k_1-k_2)^2+M^2]^\alpha[k_2^2+m^2]^\beta}
\right|_{p^2=-m^2},
\label{J012}
\end{equation}
where $n=4-2\varepsilon$ is the dimensionality of space time
(see Fig.~\ref{j012}).
Such a diagram contributes to the pole mass of the top quark \cite{JK04}.
The hypergeometric representation of this diagrams was presented in Eq.~(3.12)
of Ref.~\cite{JK04}. 
We reproduce it here for completeness: 
\begin{eqnarray}
\lefteqn{
J_{012}(\sigma,\beta,\alpha) 
=
(M^2)^{n-\sigma-\alpha-\beta}
\frac{\Gamma(\tfrac{n}{2} \!-\! \sigma)}
    {\Gamma(\sigma) \Gamma(\alpha) \Gamma(\beta)
\Gamma(\tfrac{n}{2})}
\Biggl[
\Gamma\left(\tfrac{n}{2} \!-\! \beta \right)
\Gamma\left(\alpha \!+\! \beta \!+\! \sigma \!-\! n \right)}
\nonumber \\ 
&&{}\times \Gamma\left(\beta \!+\! \sigma \!-\! \tfrac{n}{2} \right)
{}_{4}F_3 \left(\begin{array}{c|}
\alpha \!+\! \beta \!+\! \sigma \!-\! n,
\beta \!+\! \sigma \!-\! \tfrac{n}{2},
\tfrac{\beta}{2}, \tfrac{1+\beta}{2} \\
1 \!+\! \beta \!-\! \tfrac{n}{2},
\beta, \tfrac{n}{2}
\end{array} ~\frac{4m^2}{M^2} \right)
+ \left( \frac{m^2}{M^2} \right)^{n/2-\beta}
\nonumber \\
&&{}\times
\Gamma\left(\beta \!-\! \tfrac{n}{2}\right)
\Gamma(\sigma)
\Gamma\left(\alpha \!+\! \sigma \!-\! \tfrac{n}{2}\right)
{}_{4}F_3 \left(\begin{array}{c|}
\sigma,
\alpha \!+\! \sigma \!-\! \tfrac{n}{2},
\tfrac{n-\beta}{2}, \tfrac{1+n-\beta}{2} \\
1 \!+\! \tfrac{n}{2} \!-\! \beta,
n \!-\! \beta, \tfrac{n}{2}
\end{array} ~\frac{4m^2}{M^2} \right)
\Biggr] \;. 
\label{J012:hyper}
\end{eqnarray}
In the framework of differential reduction \cite{BKK}, the first hypergeometric
function in Eq.~(\ref{J012:hyper}) is expressible in terms of a Gauss
hypergeometric function and a rational function $R_1(z)$ of $z=4m^2/M^2$,
whereas the second one is expressible in terms of a ${}_3F_2$ function with one
unit upper parameter and a rational function $R_2(z)$.
Schematically, we have 
\begin{eqnarray}
&& 
{}_{4}F_3 \left(\begin{array}{c|}
\alpha \!+\! \beta \!+\! \sigma \!-\! n,
\beta \!+\! \sigma \!-\! \tfrac{n}{2},
\tfrac{\beta}{2}, \tfrac{1+\beta}{2} \\
1 \!+\! \beta \!-\! \tfrac{n}{2},
\beta, \tfrac{n}{2}
\end{array} ~z \right)
\to 
(1, \theta)
\times
{}_{2}F_1 \left(\begin{array}{c|}
I_1 \!-\! n,
\tfrac{1}{2} \!+\! I_2 \\
\tfrac{n}{2} \!+\! I_3 
\end{array} ~z \right) \!+\! R_1(z)\;,
\nonumber\\
&& 
{}_{4}F_3 \left(\begin{array}{c|}
\sigma,
\alpha \!+\! \sigma \!-\! \tfrac{n}{2},
\tfrac{n-\beta}{2}, \tfrac{1+n-\beta}{2} \\
1 \!+\! \tfrac{n}{2} \!-\! \beta,
n \!-\! \beta, \tfrac{n}{2}
\end{array} ~z \right)
\to 
(1, \theta)
\times
{}_{3}F_2 \left(\begin{array}{c|}
1,
I_1 \!-\! \tfrac{n}{2},
\tfrac{n}{2} \!+\! \tfrac{1}{2} \!+\! I_2 \\
n \!+\! I_3, 
\tfrac{n}{2} \!+\! I_4
\end{array} ~z \right) \!+\! R_2(z) \;,
\nonumber\\
&&
\end{eqnarray}
where $\theta = zd/dz$, the short-hand notation $(1, \theta)$ stands for
$(P_1(z) + P_2(z)\theta)$, with $P_i$ being rational functions (see Eqs.~(17)
and (20) in Ref.~\cite{BKK}), and $I_i$ ($i=1,\ldots,4$) are integers.
According to our criteria, there are two master integrals for this diagram that
are not expressible in terms of $\Gamma$ functions. 
However, as shown by Tarasov in Ref.~\cite{Tarasov}, solving the standard
IBP relations \cite{ibp} yields three master integrals of this
type, namely $J_{012}(1,1,1)$, $J_{012}(1,2,1)$, and $J_{012}(1,1,2)$, which is
confirmed with the help of the computer packages of
Refs.~\cite{Laporta:computer,Mertig:1998vk}.
Consequently, either our criteria are wrong or there exists an algebraic
relation between the integrals $J_{012}(1,1,1)$, $J_{012}(1,2,1)$, and 
$J_{012}(1,1,2)$, possibly including some algebraic functions depending on $z$
and products of $\Gamma$ functions.  

To find this relation, let us explore Eq.~(\ref{J012:hyper}) and present the 
master integrals in the following form: 
\begin{equation}
X_{111} =  A x_1 + B y_1 \;, 
\quad 
X_{121} = A x_2 + B y_2 \;, 
\quad 
X_{112} =  A x_3 + B y_3 \;, 
\end{equation}
where 
\begin{eqnarray}
X_{\sigma\beta\alpha} & = & (M^2)^{4-n}\left( \frac{n}{2} - 1 \right)
 J_{012}(\sigma,\beta,\alpha) \;, 
\\
A & = &  
\Gamma\left( \frac{n}{2} \!-\! 1 \right)  
\Gamma\left( 3 \!-\! n \right)  
\Gamma\left( 2 \!-\! \frac{n}{2} \right)  \;, 
\quad 
B  =   
\left( 
\frac{z}{4}
\right)^{n/2-1} 
\Gamma\left( 1 \!-\! \frac{n}{2} \right)  
\Gamma\left( 2 \!-\! \frac{n}{2} \right)  \;, 
\nonumber\\
x_1 & = & {}_2F_1\left( \tfrac{1}{2},a;b;z \right)  \;, 
\quad
x_2  =  \frac{a-1}{z} \left[ {}_2F_1\left( \tfrac{1}{2},a;b;z \right) - 1 \right] \;, 
\quad
\nonumber \\ 
x_3 &  = &  a {}_2F_1\left( \tfrac{1}{2},1+a;b;z \right)  \;, 
\quad 
y_1  =  {}_3F_2\left(1,  \tfrac{n-1}{2},c; \tfrac{n}{2}, d;z \right)  \;, 
\nonumber\\
y_2 & =& - \frac{2}{z} (d-1) {}_3F_2\left(1,\tfrac{n-1}{2},c; \tfrac{n}{2}, d-1; z \right)  \;, 
\quad
y_3   =   c {}_3F_2\left(1,\tfrac{n-1}{2},1+c; \tfrac{n}{2}, d;z \right)  \;, 
\nonumber
\end{eqnarray}
with 
\begin{eqnarray}
a  =  3 - n \;, 
\quad 
b = \frac{n}{2} \;,
\quad 
c =  2 - \frac{n}{2} \;,
\quad 
d = n \!-\! 1 \;, 
\end{eqnarray}
and 
$
{}_pF_{p-1}(\vec{a};\vec{b};z) 
$
being a hypergeometric function.
Notice that the ${}_4F_3$ functions in Eq.~\ref{J012:hyper} collapse to
${}_3F_2$ and ${}_2F_1$ functions according to Criterion~I defined in
Section~2.4 of Ref.~\cite{BKK} because upper indices exceed lower ones by
integers.
Using the relations 
\begin{eqnarray}
a\, {}_pF_q(a+1,\vec{A};\vec{B};z) & = & \left( \theta + a \right) {}_pF_q(a,\vec{A};\vec{B};z) \;, 
\nonumber  \\
(b-1)\, {}_pF_q(\vec{A};b-1,\vec{B};z) & = & \left( \theta + b - 1 \right) {}_pF_q(\vec{A}; b,\vec{B};z) \;, 
\end{eqnarray}
it is easy to obtain
\begin{eqnarray}
(3n-8) x_1 + z x_2  + 2 x_3 &=& n-2 \;,
\nonumber \\
(3n-8) y_1 + z y_2  + 2 y_3 &=& 0 \;.
\end{eqnarray}
This is equivalent to the following relation between master integrals:
\begin{eqnarray}
\lefteqn{
(3n \!-\! 8)J_{012}(1,1,1)
+  4 m^2 J_{012}(1,2,1) 
+ 2 M^2 J_{012}(1,1,2)}  
\nonumber\\
&=& 
2 (M^2)^{n-3} 
\Gamma\left( \frac{n}{2} \!-\! 1 \right)  
\Gamma\left( 3 \!-\! n \right)  
\Gamma\left( 2 \!-\! \frac{n}{2} \right)  \;. 
\label{main}
\end{eqnarray}
Eq.~(\ref{main}) does not only represent a relation between Feynman diagrams,
but also a relation between hypergeometric functions ${}_pF_q$, which is newly
derived from the differential reduction technique.

For $m^2=M^2$, we have
\begin{equation}
(3n - 8)J_{011}(1,1,1)+6 m^2 J_{011}(1,1,2) 
= 2 (m^2)^{n-3} 
\Gamma\left( \frac{n}{2} \!-\! 1 \right)  
\Gamma\left( 3 \!-\! n \right)  
\Gamma\left( 2 \!-\! \frac{n}{2} \right)  \;. 
\label{aux}
\end{equation}
The integrals 
$J_{011}(1,1,1)$ 
and 
$J_{011}(1,1,2)$ with $m^2=M^2$ 
are master integrals of the package {\tt ON-SHELL2} \cite{onshell2}. 
Eq.~(\ref{aux}) coincides with Eq.~(4.45) of Ref.~\cite{DK01}, where it was
derived by studying the analytical coefficients of the higher-order 
$\ep$ expansion of diagram $J_{011}(1,2,2)$ performed in
Ref.~\cite{basis}.

According to Ref.~\cite{czakon}, the last terms in Eqs.~(\ref{main}) and
(\ref{aux}) may be identified with a two-loop bubble diagram with one massive
line divided by $M^2(n-2)$, so that Eq.~(\ref{main}) is an algebraic
relation between four two-loop diagrams. 
However, that two-loop bubble diagram, having two massless lines, cannot be
obtained from the original two-loop sunset diagram with two massive lines by
the contraction of lines.
Consequently, Eq.~(\ref{main}) cannot be derived from standard IBP relations
for two-loop sunset diagrams.

At this point, a comment on the structural difference between the
differential-reduction and IBP approaches seems appropriate.
Eq.~(\ref{main}) represents a linear relation between master integrals that
contains an inhomogenious term involving $\Gamma$ functions on the right-hand
side.
However, it is clear from the structure of the IBP relations that the IBP
method can only result in relations between integrals in which the prefactors
are rational functions, while $\Gamma$ functions do not appear in IBP
relations.
This is another reason why Eq.~(\ref{main}) cannot be derived using the IBP
method.

In conclusion, counting the number of irreducible master integrals for the
diagram defined by Eq.~(\ref{J012}) using our criteria on the one hand and
the explicit analytical solution \cite{Tarasov,Laporta} of the standard
integration-by-part relations \cite{ibp}, as implemented in the widely used
program packages \cite{Laporta:computer,Mertig:1998vk}, on the other hand, we
encountered a mismatch. 
This motivated us to established a new algebraic relation, namely
Eq.~(\ref{main}).
This is a first example where our criteria allow us to expose a new relation 
between master integrals of the standard integration-by-part technique.

Finally, we wish to mention that the final result of Ref.~\cite{JK04} does not
depend on the number of master integrals.
In fact, all master integrals are expressible in terms of hypergeometric
functions, and the analytical coefficients of the $\ep$ expansions of the
latter were constructed in Ref.~\cite{expansion}.

\vspace{5mm}
\noindent 
{\bf Acknowledgments} \\
We are grateful to P.A.~Baikov, A.I.~Davydychev, A.G.~Grozin, A.V.~Kotikov,
and O.L.~Veretin for their interest in our work and for cross checking
Eq.~(\ref{main}).
We thank P.A.~Baikov and A.G.~Grozin for sharing with us their observation that
the propagator diagrams entering Eq.~(\ref{main}) may be considered as special
cases of more general propagator diagrams with five internal lines.
This work was supported in part by the German Federal Ministry for Education
and Research BMBF through Grant No.\ 05~HT6GUA, by the German Research
Foundation DFG through the Collaborative Research Centre No.~676
{\it Particles, Strings and the Early Universe---The structure of Matter and
Space Time}, and by the Helmholtz Association HGF through the Helmholtz
Alliance Ha~101 {\it Physics at the Terascale}.

\end{document}